\documentclass[11pt,twoside]{article} 
\usepackage{acta-info,euler}
\usepackage{graphicx,gensymb}
\usepackage{epstopdf}
\newcommand{\vol}{\textbf{2,} 1 (2010) 90--98}   
\newcommand{\amss}{\amssubj{%
94C99, 68U05
}}     

\newcommand{\CRs}{\CRsubj{%
B.7.2
}}   
\newcommand{\keyw}{\keywords{%
chaos, Chua's circuit, singularity, trajectory, 3D visualization
}}

\begin{document}

\title{%
Macro and micro view on~steady states in~state space
}

\maketitle




\twoauthors{%
\href{http://kpi.fei.tuke.sk}{Branislav Sobota}
}{%
\href{http://kpi.fei.tuke.sk}{%
Dept. of Computers and Informatics\\
FEEI, Technical University of Ko\v{s}ice\\
Letn\'{a} 9, 04200 Ko\v{s}ice, Slovakia}
}{%
\href{mailto:Branislav.Sobota@tuke.sk}{Branislav.Sobota@tuke.sk}
}{%
\href{http://kteem.fei.tuke.sk}{Milan Guzan}
}{%
\href{http://kteem.fei.tuke.sk}{%
Dept. of Theoretical Electrotechnics\\ and Electrical Measurement\\
FEEI, Technical University of Ko\v{s}ice\\
Letn\'{a} 9, 04200 Ko\v{s}ice, Slovakia}
}{%
\href{mailto:Milan.Guzan@tuke.sk}{Milan.Guzan@tuke.sk}
}



\short{%
B. Sobota, M. Guzan
}{%
Macro and micro view on steady states in state space
}

\begin{abstract}
This paper describes visualization of chaotic attractor and elements of the singularities in 3D space. 3D view of these effects enables to create a demonstrative projection about relations of chaos generated by physical circuit, the Chua's circuit. Via macro views on chaotic attractor is obtained not only visual space illustration of representative point motion in state space, but also its relation to planes of singularity elements. Our created program enables view on chaotic attractor both in 2D and 3D space together with plane objects visualization -- elements of singularities.
\end{abstract}

\section{Introduction}
The visualization is good idea to show imagines, ideas, design, construction, realization or effects. It is also one way of verification before realization our goals. Computer based visualization brings utilization of physical, or simulated electric parameters course graphical interpretation in non-linear circuit theory together with other fields. From beginning it was used 2D visualization with possibility of color utilization to be more illustrative or to explain actions proceeding in non-linear circuits \cite{4,2,3,1}. High-performance or parallel computers enable to take advantage of 3D state space axonometry \cite{5}. Actual available solutions provide high performance visualization suitable for 3D interactive presentation of processes and effects with support for over million saturated colors in hi-resolution mode and for use in all graphics-intensive applications. This paper describes actual possibilities of PC for visualization of steady states of chaos generating circuit.

\section{Methods used for trajectory visualization}
In last 24 years there was intensive interest of scientific community to analyse and applied Chua's circuit generating chaos. Presentation of trajectories needs to solve system (\ref{eq1}) describing physical Chua's circuit. 
\begin{eqnarray}\label{eq1}
C_1 ( du_1/dt ) =& G (u_2 - u_1 ) - g (u_1 ) - I &= Q_1 \nonumber\\
C_2 ( du_2/dt ) =&     G (u_1 - u_2 ) + i        &= Q_2 \\\nonumber
 L ( di/dt ) =&         -u_2 - \varrho i         &= Q_3
\end{eqnarray}
where
\begin{eqnarray}\label{eq2}
g (u_1) = & m_2 u_1 + 1/2(m_1 - m_0)(|u_1 - B_P| - |u_1 + B_P|) + \nonumber\\
 & + 1/2(m_2 - m_1)(|u_1 - B_0| - |u_1 + B_0|)
\end{eqnarray}

Next we consider control pulse $I=0$, the resistance of the inductance $\varrho=0$.

\begin{figure}[h]
\begin{center}
\includegraphics[scale=0.8, width=.99\textwidth]{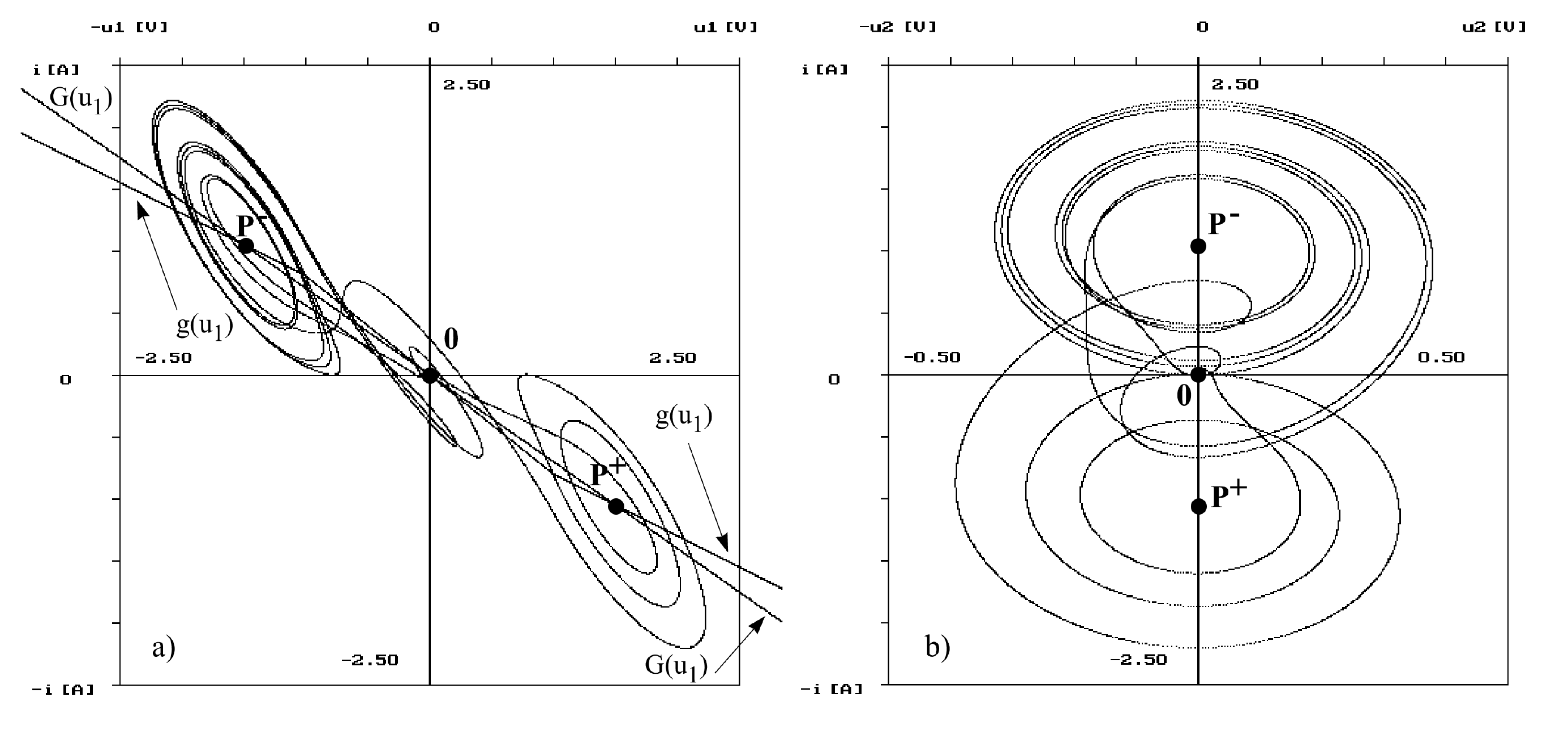}
\end{center}
\caption{Monge projection of chaotic attractor system (\ref{eq1}), for parameters (\ref{eq3}) to plane: a) $i-u_1$, b) $i-u_2$}
\label{fig1}
\end{figure}

For parameters (\ref{eq3}) in \cite{6} there were found chaotic attractors showed in Fig.~\ref{fig1} in Monge projection. 
\begin{eqnarray}\label{eq3}
C_1=1/9, C_2=1, L=0.142857, G=0.7, \nonumber\\
m_0=-0.8, m_1=-0.5 , m_2=5 , B_p=1, B_0=14
\end{eqnarray}

Computer program was designed in C language by author of \cite{7} and used for clarifying of place in state space where chaos originates \cite{8}. It is only short segment of intersection two surfaces related to circuit singularities $P+$ and $0$, or $0$ and $P-$. In despite of explaining with help of tables and 2D presentation was definite, 3D visualization provides faster and lighter illustration of actions which proceed in specific non-linear circuit. Therefore 3D visualization is valued as from scientific as from edifying point of view \cite{9}.

\section{Visualization in 3-dimensional space}

Chaos visualizing system was designed for visualization Chua's attractor in 3D space in real time and it is based on visualizing kernel developed on DCI FEEI TU Ko\v{s}ice \cite{10}. An application is implemented in C++ language using OpenGL graphics library. The application can work with \emph{I-V characteristics}, Chua's attractor trajectory, or limit cycle and it can visualize elements of the singularity planes.  Additionally this visualization depicts representative point movement and it creates chaotic attractor using two basic modes (\emph{continuous} and \emph{sequential}). This application can be used not only for concrete Chua's circuit. It is usable also for Chua's circuit like structures analyzed in~\cite{12,11}.

Chaos visualizing system provides three basic visualizing modes: \emph{continuous mode, sequential mode} and \emph{I-V characteristics visualization mode}. System allows using four projection types (\emph{3D $u_1 - i - u_2$ projection} (see Fig.~\ref{fig5}), \emph{2D $i - u_2$ projection, 2D $i - u_1$ projection} and \emph{2D $u_2 - u_1$ projection}) for better-examined circuit understanding. Settable basic visualizing parameters for chaotic attractor visualization are: \emph{drawing speed, points omission, chaotic attractor point size, comet length} and \emph{attractor colour}. The combination of these parameters defines final visualization form of chaotic attractor. In case of 3D graphic accelerator supported acceleration of graphic interface OpenGL is drawing of 3D primitives accelerated by graphic card. It enables increasing performance and using some graphical improvements cannot be used without acceleration in real time. Second way is use of parallel computational system~\cite{13,14} for faster or better visualization.

\subsection{Visualization program utilization}
\begin{figure}[hbt]
\begin{center}
\includegraphics[width=.48\textwidth]{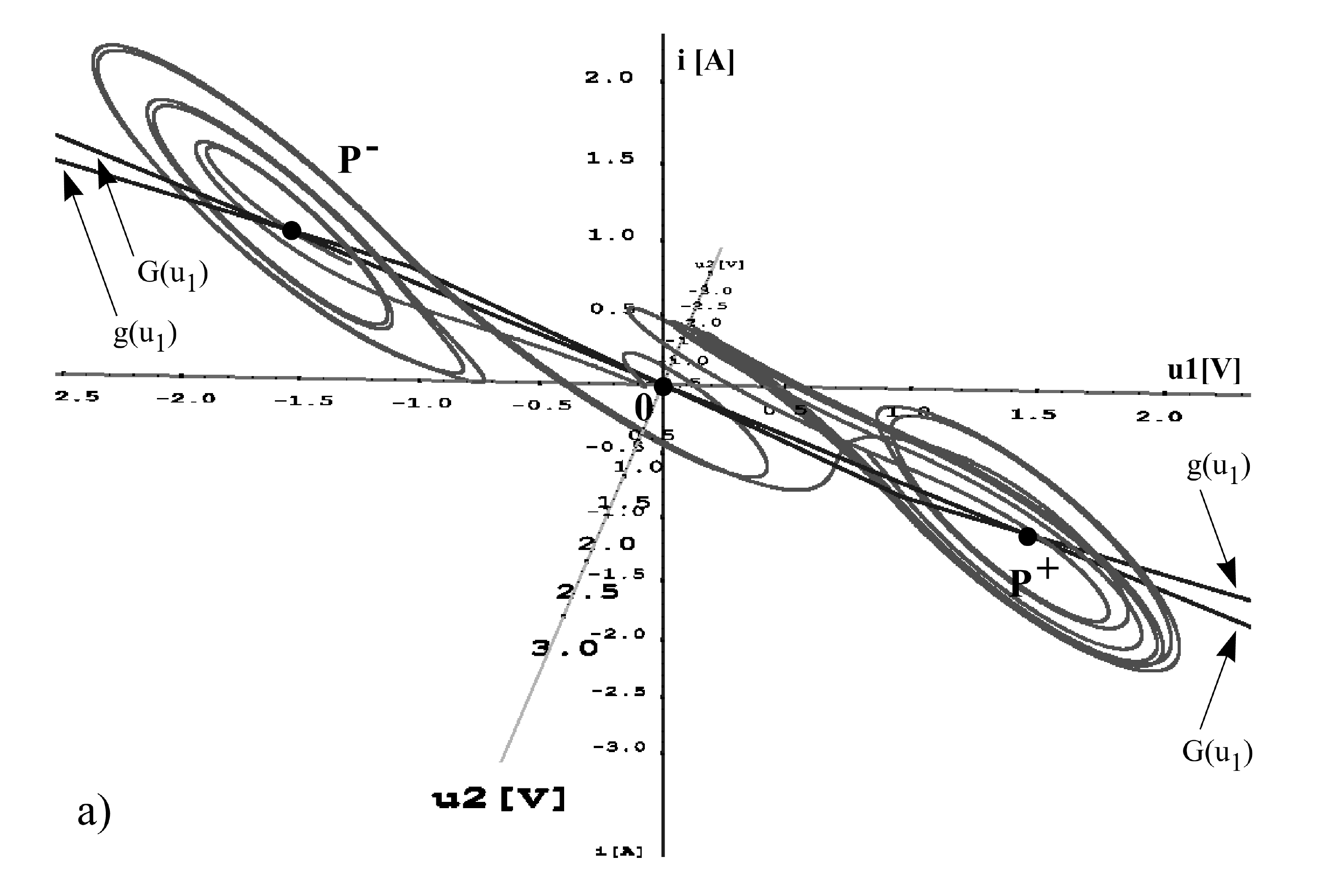}
\includegraphics[width=.48\textwidth]{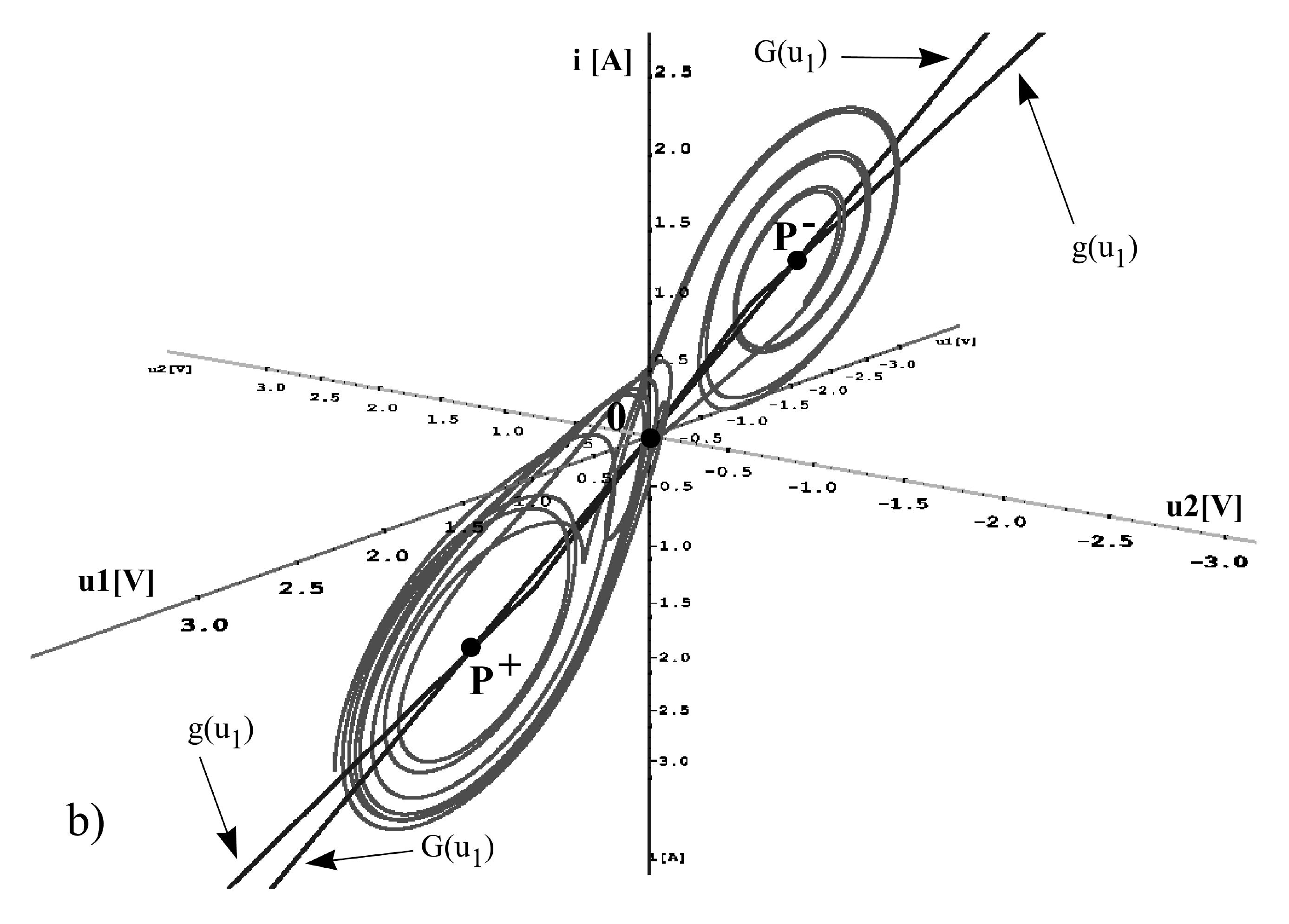}
\includegraphics[width=.48\textwidth]{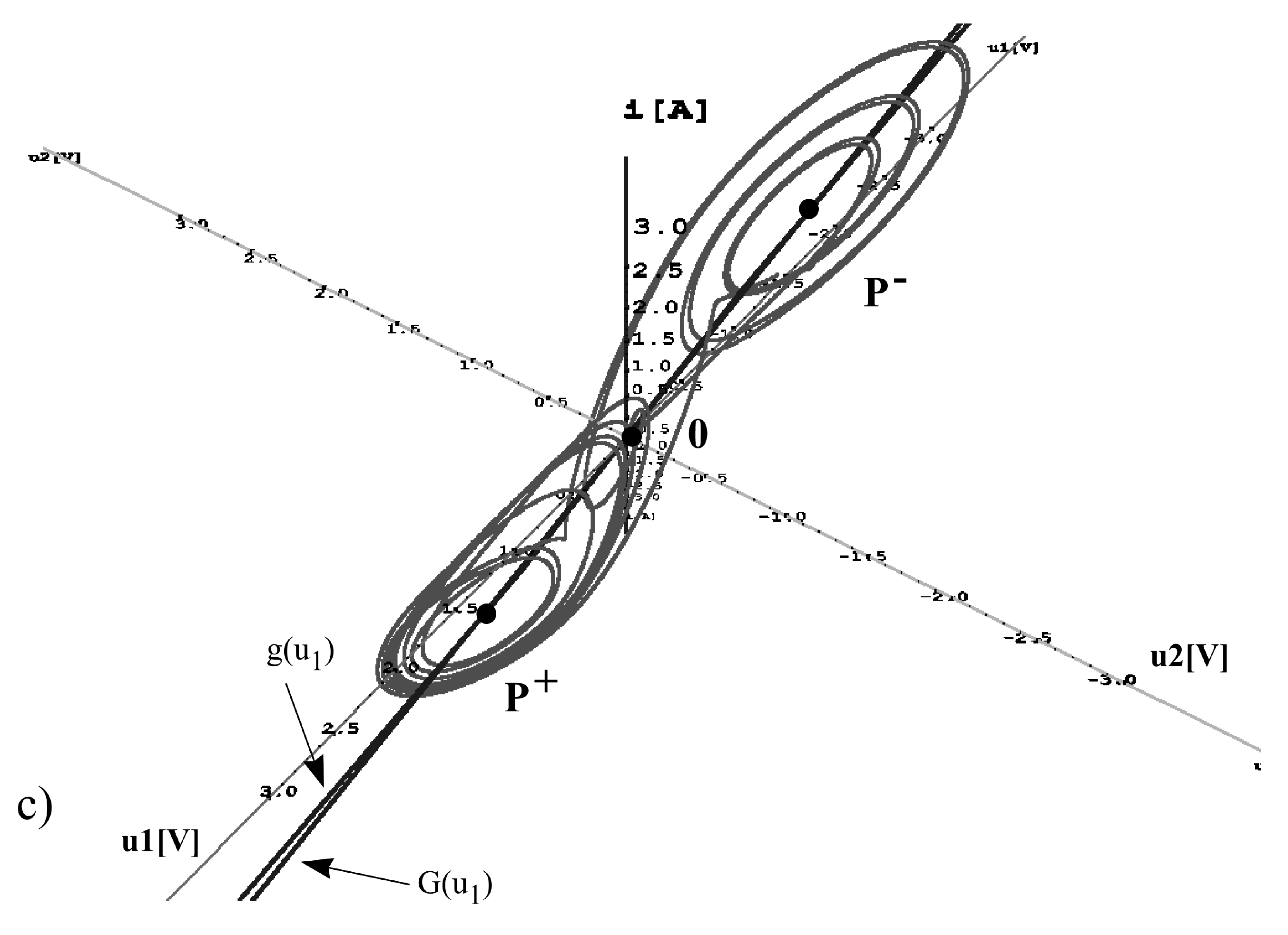}
\includegraphics[width=.48\textwidth]{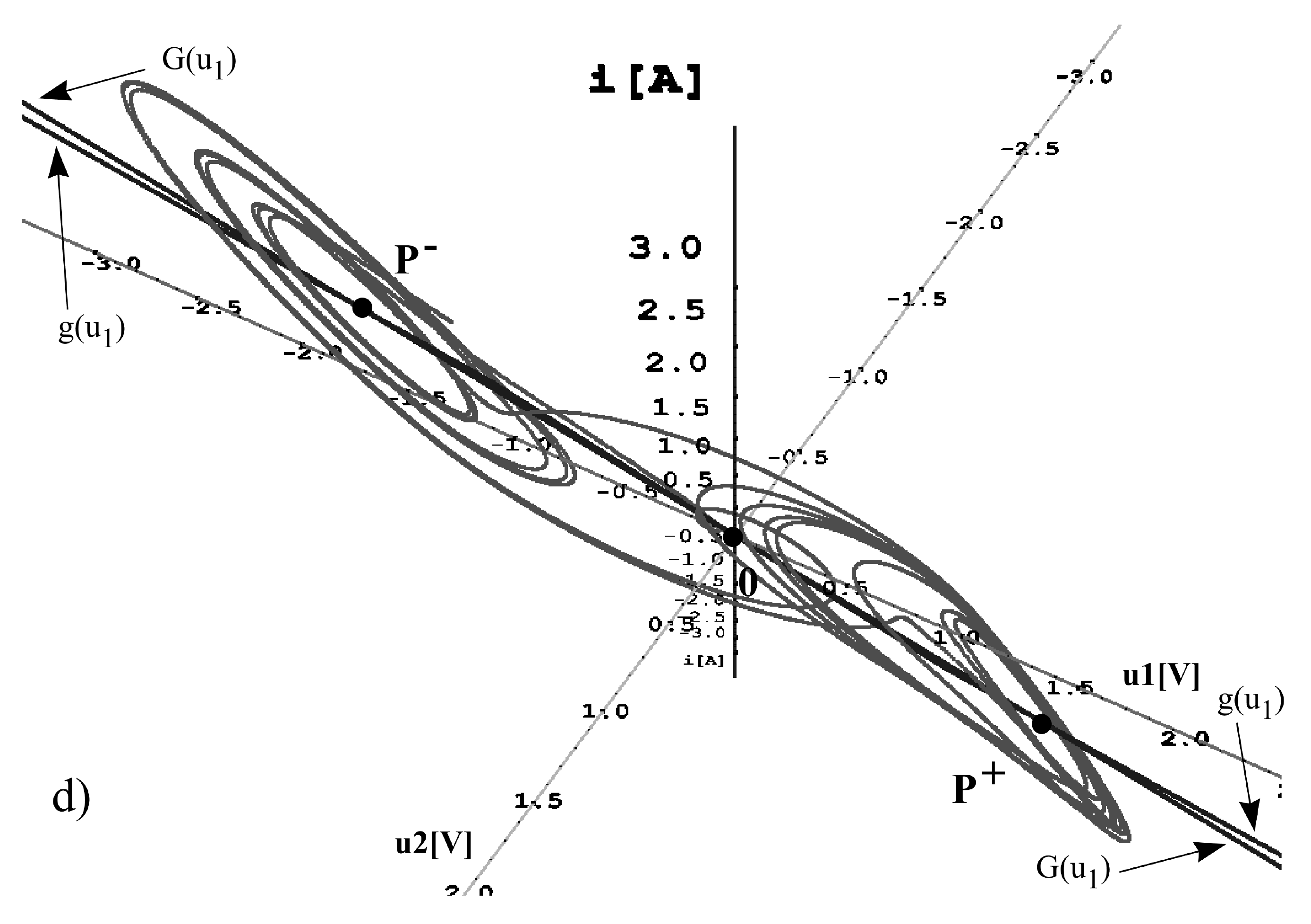}
\end{center}
\caption{Chaotic attractor depicted in continuous mode}
\label{fig2}
\end{figure}

3D visualization of chaotic attractor for parameters (\ref{eq3}) is shown in Fig.~\ref{fig2}. To next manipulation with chaotic attractor as 3D object is necessary to fulfill the following steps:

\begin{itemize}\addtolength{\itemsep}{-0.6\baselineskip}
\item Load input file -- chaotic attractor. Input file size is above 500 MB, therefore program enables to choose number of trajectory points, which is loaded from file and consequently displayed.
\item Load \emph{I-V characteristics} $G(u1)$ and $g(u1)$ from files.
\item Set background color for scene displaying and
\item Set colors and line-width of \emph{I-V characteristics} and \emph{chaotic attractor}.
\end{itemize}
Views on chaotic attractor from various sides can be obtained by rotation of camera position horizontally and also vertically around visualized object. Fig.~\ref{fig2}a and b show horizontal camera swing out, while Fig.~\ref{fig2}c, d show vertical camera swing out to chaotic attractor. In this way top view on observed object can be obtained.
\begin{figure}[hbt]
\begin{center}
\includegraphics[width=.48\textwidth]{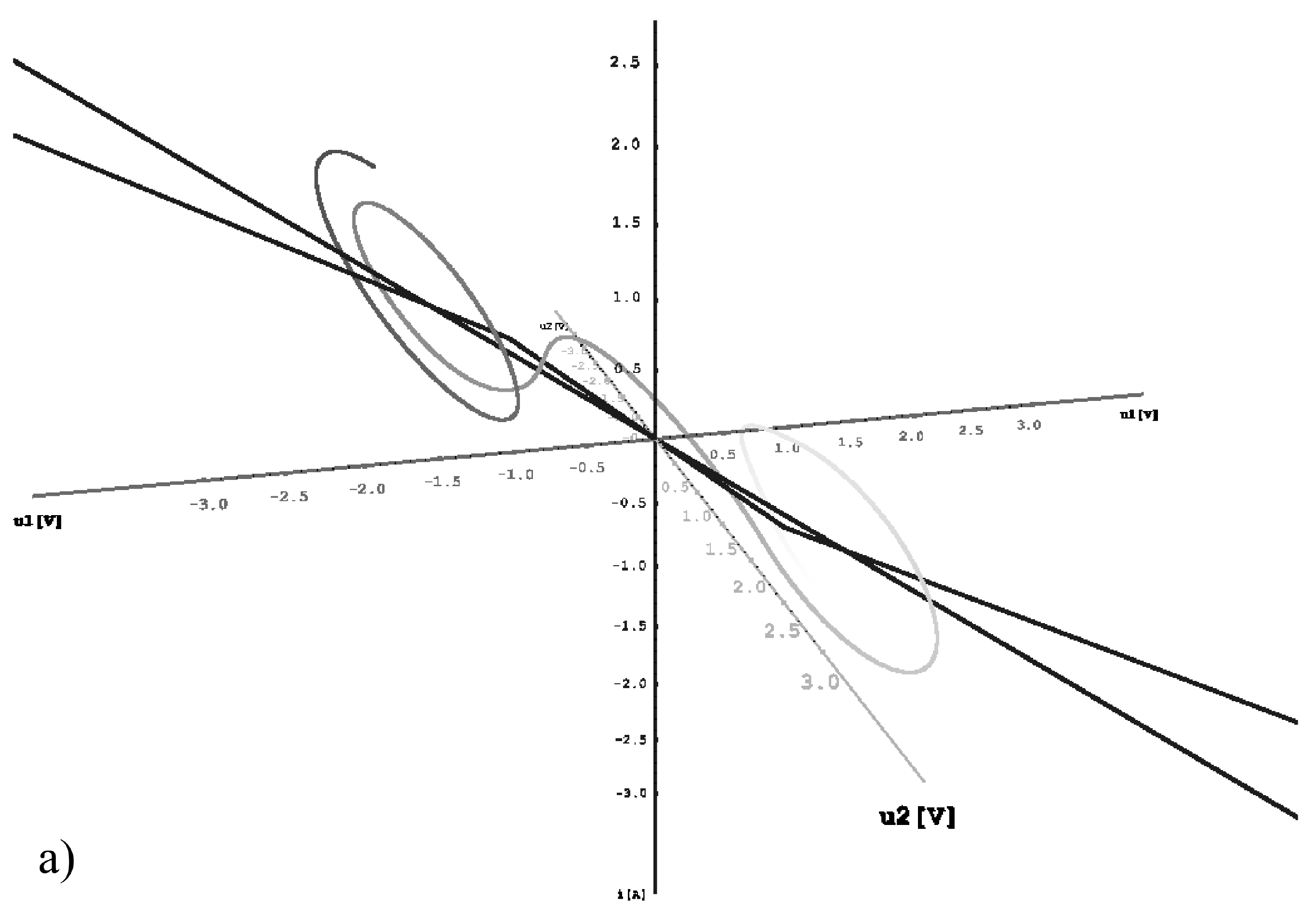}
\includegraphics[width=.48\textwidth]{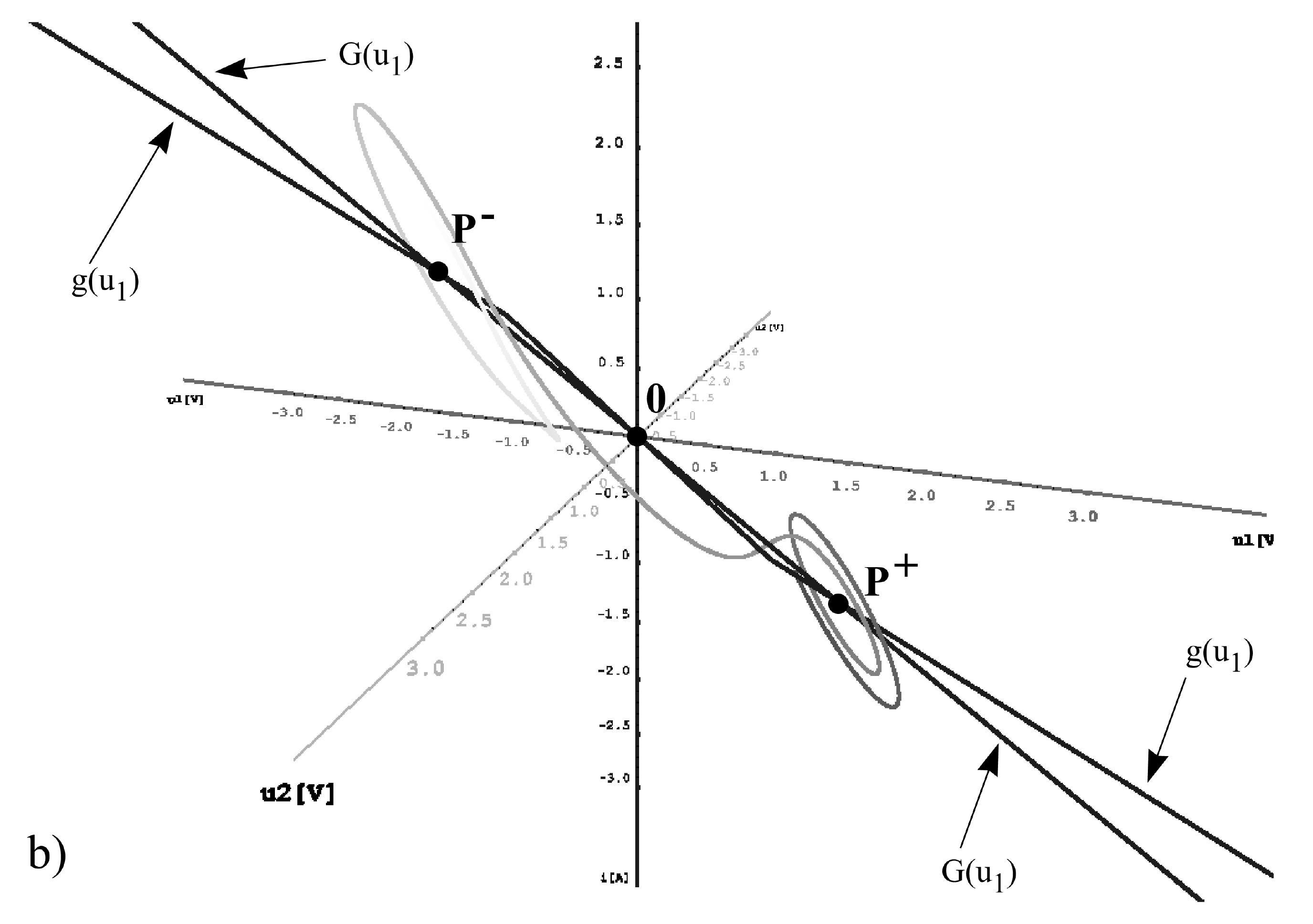}
\includegraphics[width=.48\textwidth]{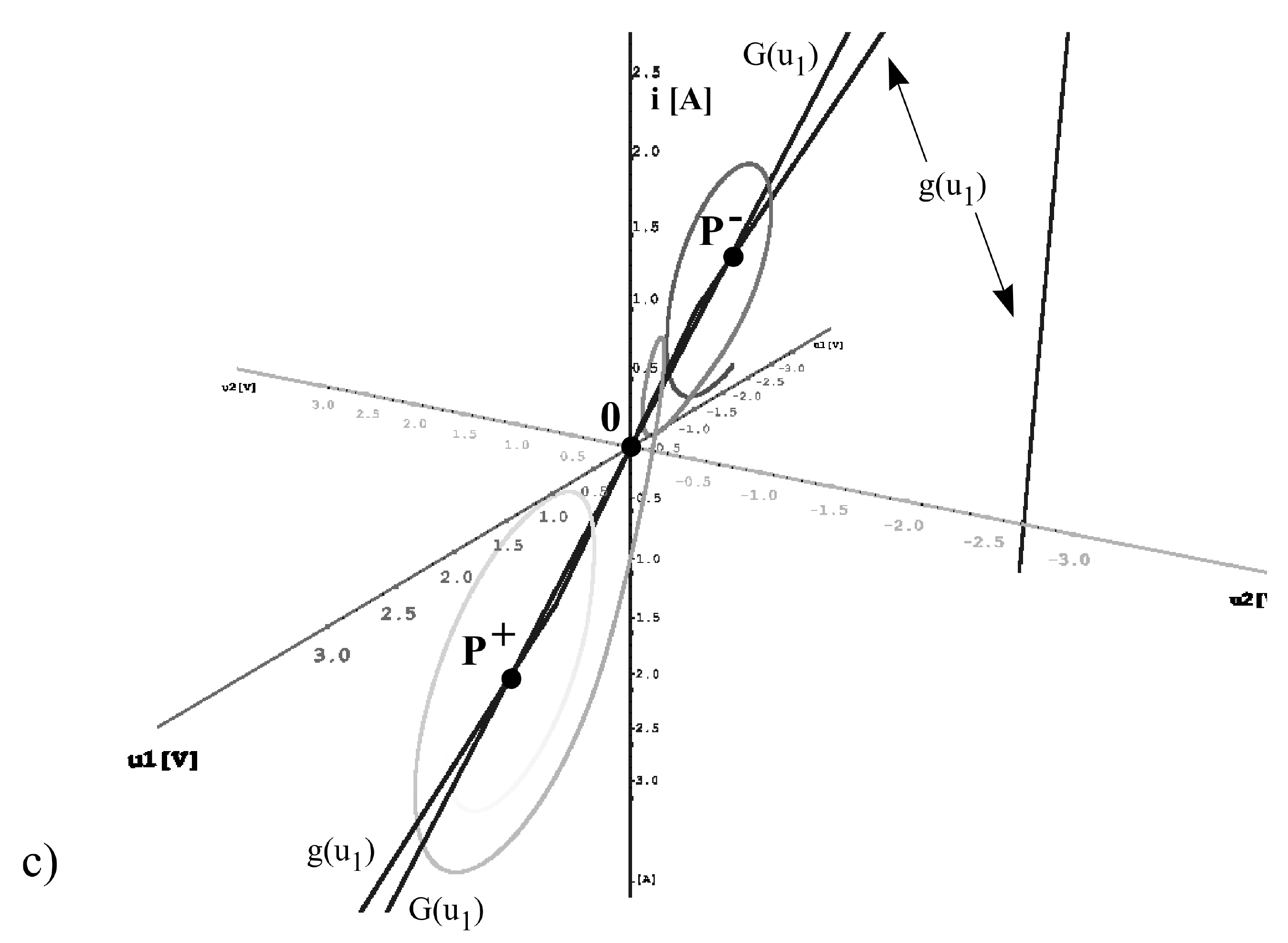}
\end{center}
\caption{Chaotic attractor depicted in sequential mode with comet effect}
\label{fig3}
\end{figure}
On Fig.~\ref{fig3} are displayed different looks to the same chaotic attractor. It is visualizing in sequential mode with comet effect \cite{15}. The comet length is adjustable from 512 to 16384 bits. Fig.~\ref{fig2} and also Fig.~\ref{fig3} show that left and right discs of chaotic attractor are situated in plane. It is possible to display this plane in the program. Mathematical description of this plane presented in \cite{8} is outlined by the following equation:
\begin{equation}\label{eq4}
y_1 = \alpha_{11}\Delta u_1 + \alpha_{12}\Delta u_2 + \alpha_{13}\Delta i. 
\end{equation}
It is available to define in application menu input parameters for appropriate planes as e.g.: singularity coordinates ($i, u_1, u_2$), eigenvectors ($\alpha_{11}, \alpha_{12}, \alpha_{13}$), width, length and planes colors. The elements of the singularity planes are displayed by application using of selected colors. Fig.~\ref{fig4}a shows these elements of the singularities $EP+$ and $EP-$ representation as parallel planes getting across singularities $P+$ and $P-$. Fig.~\ref{fig4}b shows also the third singularity element $0$, called $E0$. Fig.~\ref{fig4}b shows, that plane $E0$ is not parallel with $EP+$ and $EP-$.  Via macro views on chaotic attractor showed in Fig.~\ref{fig2}--~\ref{fig4} we obtain not only visual space illustration of representative point motion in state space, but also its relation to planes of singularity elements.
\begin{figure}[hbt]
\begin{center}
\includegraphics[width=.48\textwidth]{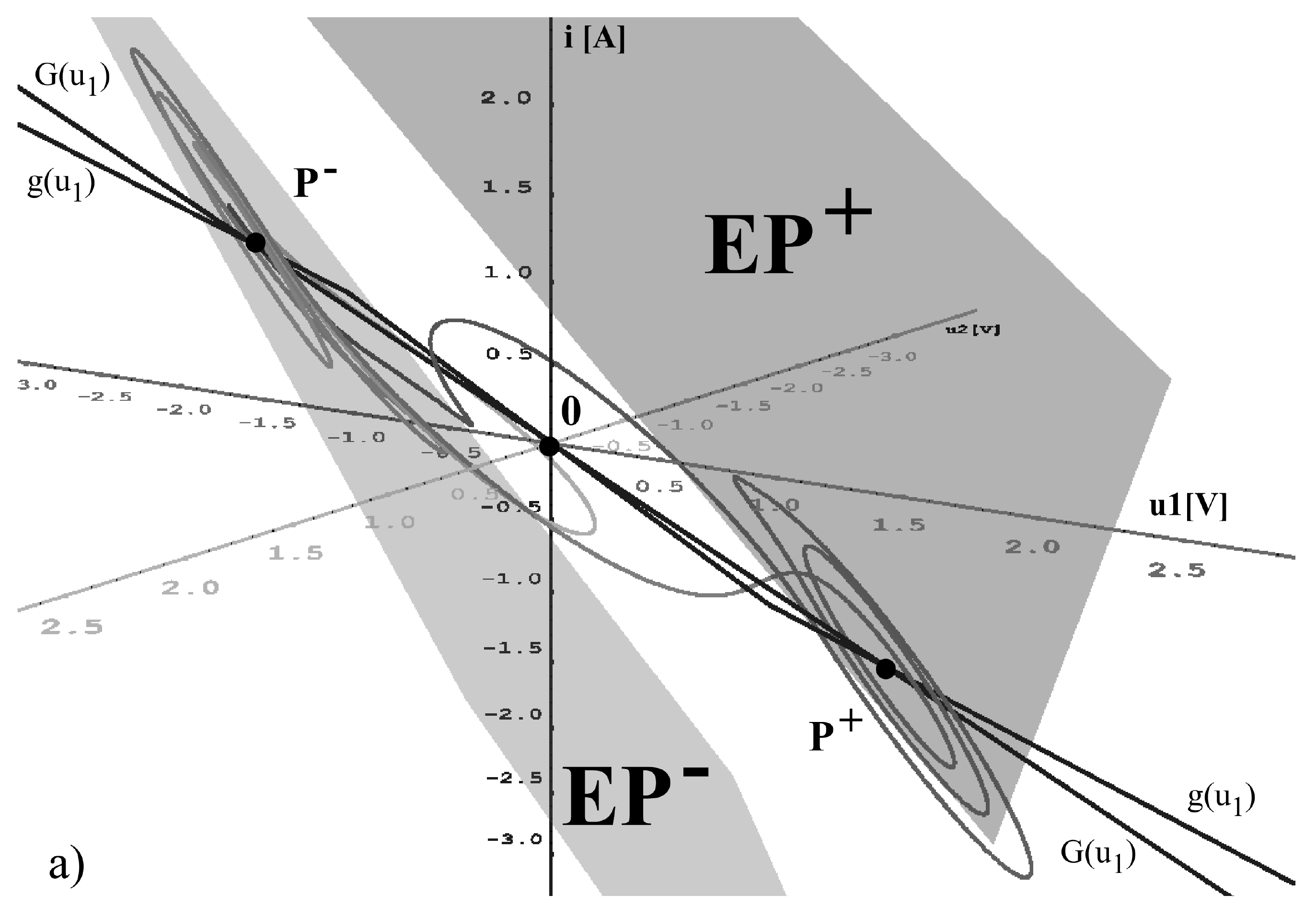}
\includegraphics[width=.48\textwidth]{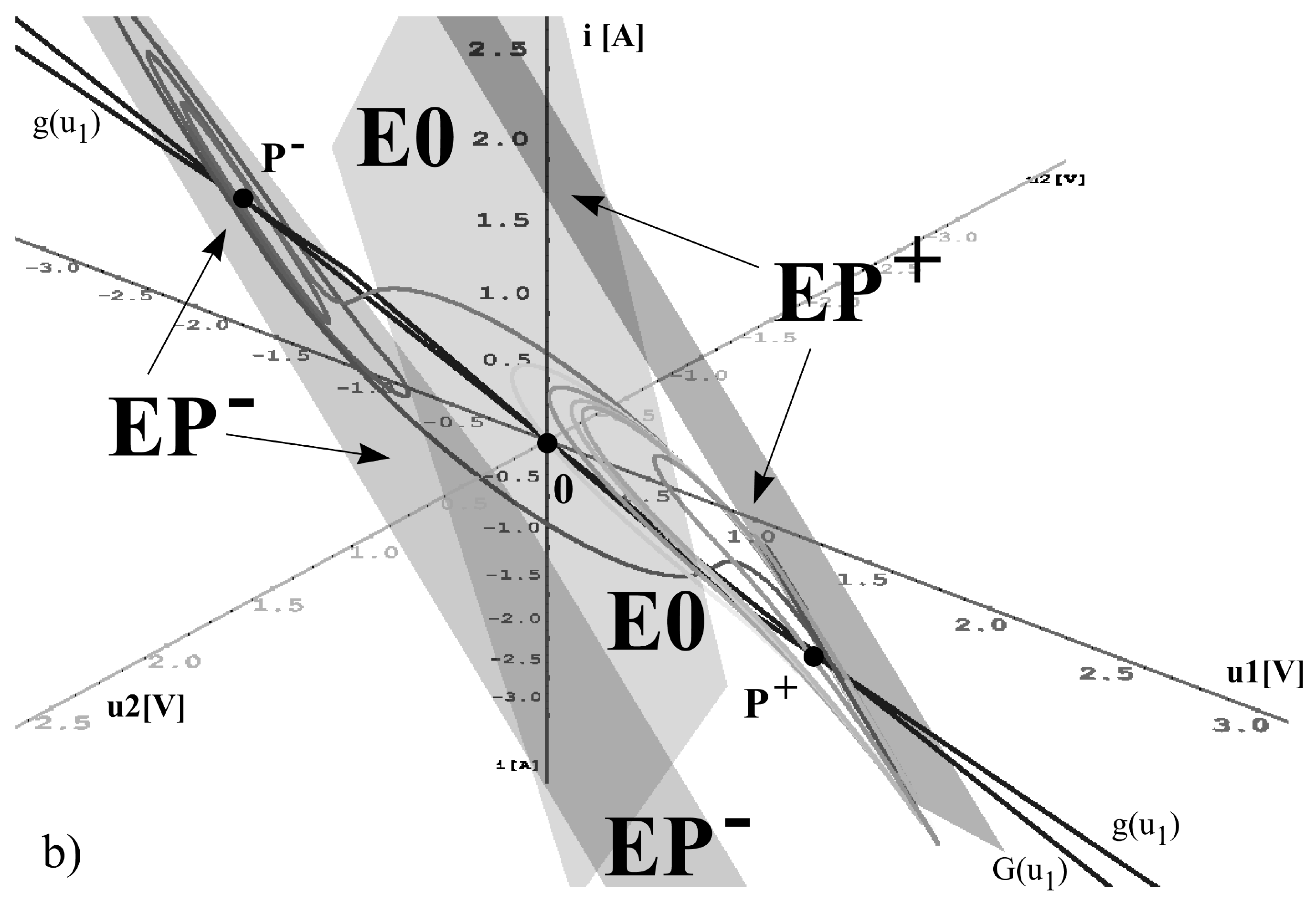}
\end{center}
\caption{Visualization of singularity elements in state space: a) $EP- $ and $EP+$ b) $E0, EP- $ and $EP+$}
\label{fig4}
\end{figure}
Vertical camera swing round with regard to Fig.~\ref{fig4} enables to see such a part of state space, where chaos arises in Chua's circuit. It is intersection of two planes $EP+$ and $E0$ or $E0$ and $EP-$. This situation is displayed in macro view on Fig.~\ref{fig5}a. Marking of conjunction region of mentioned elements $E0$ and $EP-$ enables to obtain micro view to just site of state space, where chaos originates. 
\begin{figure}[hp]
\begin{center}
\includegraphics[width=.90\textwidth]{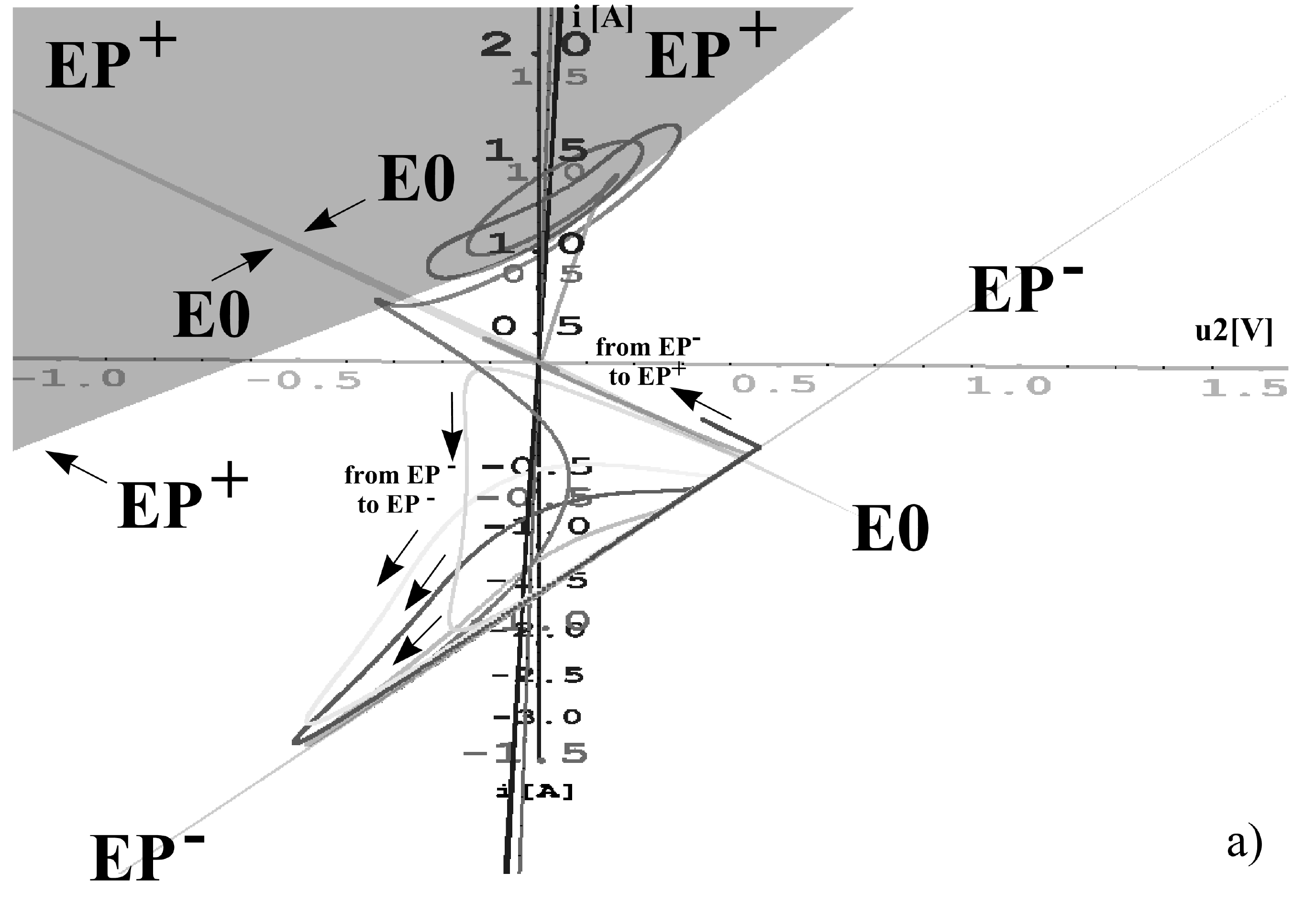}
\includegraphics[width=.90\textwidth]{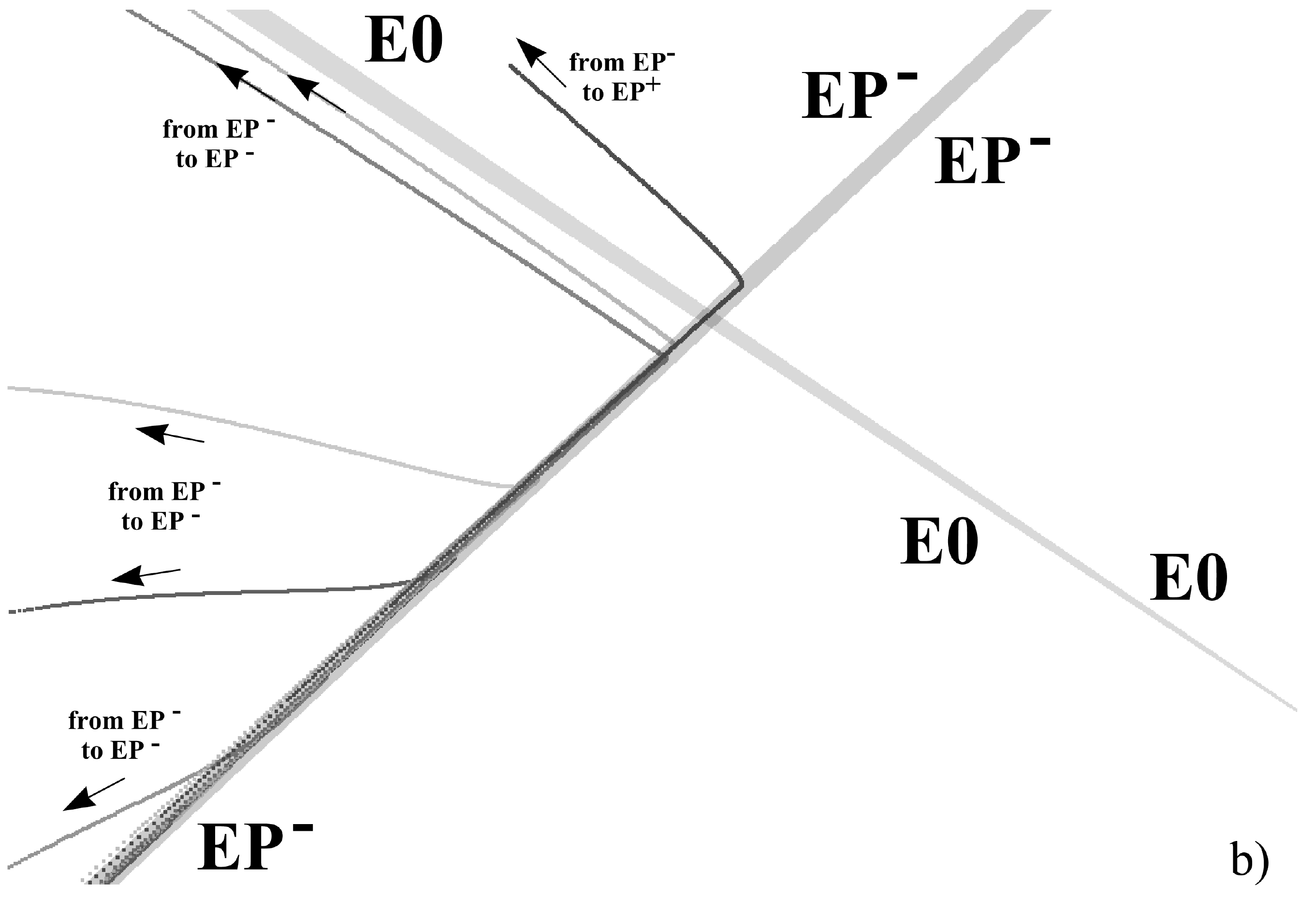}
\end{center}
\caption{3D view on intersection of singularity elements $EP- $ and $E0$ (view on place of chaos origination in Chua's circuit)}
\label{fig5}
\end{figure}
Website \href{http://kteem.fei.tuke.sk/guzan/ausi09}{http://kteem.fei.tuke.sk/guzan/ausi09} contains dynamic versions of some pictures mentioned in this article. Video appears jerky. This is caused by camera position rotation around object (usually 5\degree). Program enables to set angle step of camera position rotation. It was used when Fig.~\ref{fig5} was generated.

\section{Conclusion}
The 3D visualization brings new dimensions to the visualization of physical or electric effects. Visualization of chaotic attractor and elements of the singularities provides better understanding of representative point movement in state space, what was still possible only with help of representation in projection planes. From computer graphics point of view  are produced big data-sets. Possible parallel processing (e.g. on multi-core or multi-computers platform) shortens computational time. Big-screen display solutions increase quality and ability of immersion into 3D space. It is possible to use finer integration step for better quality and more detailed visualization output (continuous trajectory displaying) in this case. Generally, there are two important application areas in physical or electric effects visualization where big-display environments are used: displaying images at very high resolution in real time exceeding those of available screens (monitors) and/or graphics cards and providing a larger field-of-view and better immersion into the explored attractor`s space. Whole system enables visualization for user defined Chua`s attractor where user has in standard visualizing mode (3D projection) 6 degrees of freedom for motion in explored attractor`s space, which mean translation in 3 axis and rotating around them. It means that system is capable to visualize the Chua's attractor from any point and from any angle.  Actual available solution provide so unique big wide-screen high performance visualization solution suitable for 3D interactive presentation of attractor`s space. Input devices are standard keyboard and mouse generally, but space mouse or other specialized input device can be also used.

\section*{Acknowledgements}
This work is supported by VEGA grant project No. 1/0646/09: ``Tasks solution
for large graphical data processing in the environment of parallel, distributed
and network computer systems'', by KEGA no. 3/6386/08: ``E-learning and web oriented education technology of subjects from the field of electrical measurement for presentation and distance form of study''  and by Agency of the Ministry of Education of the
Slovak Republic for the Structural Funds of the EU under the project Center of Information and Communication Technologies for Knowledge Systems
(project number: 26220120020).

\rightline{\emph{Received:  July 15, 2009  {\tiny \raisebox{2pt}{$\bullet$\!}} Revised: February 28, 2010}}     

\end{document}